\documentclass[10pt,conference]{IEEEtran}
\IEEEoverridecommandlockouts

\usepackage{cite}
\usepackage{amsmath,amssymb,amsfonts}
\usepackage{graphicx}
\usepackage{textcomp}
\usepackage{xcolor}
\usepackage{microtype}
\usepackage{listings}
\usepackage{etoolbox}
\usepackage{enumitem}
\usepackage{tikz}
\usepackage{booktabs}
\usepackage{siunitx}
\usepackage{balance}
\usepackage[most]{tcolorbox}
\usepackage{xcolor}
\usepackage{makecell}

\renewcommand\theadfont{\bfseries\footnotesize}

\definecolor{rqblue}{HTML}{1F4E79}
\definecolor{rqlight}{HTML}{F4F8FB}

\newtcolorbox{rqanswerbox}[1]{
  enhanced,
  breakable,
  colback=rqlight,
  colframe=rqblue,
  boxrule=0.6pt,
  leftrule=2.2pt,
  arc=2pt,
  left=6pt,
  right=6pt,
  top=5pt,
  bottom=5pt,
  title=\textbf{#1},
  coltitle=white,
  colbacktitle=rqblue,
  attach boxed title to top left={xshift=0pt,yshift=-1.5mm},
  boxed title style={
    sharp corners,
    boxrule=0pt,
    left=5pt,
    right=5pt,
    top=2pt,
    bottom=2pt
  }
}

\usetikzlibrary{arrows.meta,positioning,shapes.geometric,fit,backgrounds,calc}
\usepackage{hyperref}

\tikzset{
    evobox/.style={rectangle, rounded corners=2pt, draw=blue!55!black,
                   fill=blue!6, align=center, font=\scriptsize, inner sep=3pt,
                   text width=30mm, minimum height=8mm},
    evohot/.style={rectangle, rounded corners=2pt, draw=orange!80!black,
                   line width=0.8pt, fill=orange!14, align=center,
                   font=\scriptsize, inner sep=3pt, text width=30mm,
                   minimum height=8mm},
    evoboxc/.style={evobox, text width=38mm, inner xsep=5pt,
                    execute at begin node={\hyphenpenalty=10000\exhyphenpenalty=10000}},
    evodec/.style={diamond, aspect=2.2, draw=blue!55!black, fill=blue!6,
                   align=center, font=\scriptsize, inner sep=1pt},
    evoflow/.style={-{Latex[length=1.8mm]}, semithick, draw=black!70}
}

\hypersetup{
    colorlinks=true,
    linkcolor=blue!50!black,
    citecolor=blue!50!black,
    urlcolor=blue!50!black
}

\definecolor{lstbg}{HTML}{F7F7F7}
\definecolor{lstkw}{HTML}{0B5394}
\definecolor{lstcmt}{HTML}{6A737D}
\definecolor{lststr}{HTML}{A31515}

\lstdefinestyle{evopbt-java}{%
    language=Java,
    backgroundcolor=\color{lstbg},
    basicstyle=\ttfamily\footnotesize,
    keywordstyle=\color{lstkw}\bfseries,
    commentstyle=\color{lstcmt}\itshape,
    stringstyle=\color{lststr},
    showstringspaces=false,
    columns=fullflexible,
    keepspaces=true,
    breaklines=true,
    breakatwhitespace=false,
    captionpos=b,
    morekeywords={Property,ForAll,Assume,assertThatThrownBy,TargetMethod,
                  JqwikProperty,PropertyChecker,PropertyRegistry,
                  ExecutionObserver,TestChromosome,DynaMOSA,beforeStatement,
                  afterStatement,testExecutionFinished,isCovered,Test,
                  assertEquals,assertTrue,assertFalse}
}
\definecolor{diffadd}{HTML}{E6FFED}
\definecolor{diffdel}{HTML}{FFEEF0}
\lstdefinestyle{evopbt-diff}{%
  style=evopbt-java,
  moredelim=[is][\colorbox{diffadd}]{+}{+},
  moredelim=[is][\colorbox{diffdel}]{-}{-}%
}

\def\BibTeX{{\rm B\kern-.05em{\sc i\kern-.025em b}\kern-.08em
    T\kern-.1667em\lower.7ex\hbox{E}\kern-.125emX}}

\newcommand{\tool}{\textsc{Progress}}

\begin{document}

\title{PROGRESS: Property-Guided Regression Search for Semantic Falsification}

\author{
\IEEEauthorblockN{Davis Tocheuk Mo\textsuperscript{1},\quad Noshin Ulfat\textsuperscript{1},\quad Matthew B. Dwyer\textsuperscript{2},\quad Soneya Binta Hossain\textsuperscript{1}}
\IEEEauthorblockA{%
\textsuperscript{1}University of Texas at Dallas, Texas, USA\\
\textsuperscript{2}University of Virginia, Virginia, USA\\[0.5ex]
\{Davis.Mo, noshin.ulfat, sbhossain\}@utdallas.edu,\quad matthewbdwyer@virginia.edu}
}

\maketitle

\begin{abstract}

Search-based regression-test generation is highly effective at exploring complex program structures and producing test suites with high structural coverage. Its \textit{test oracles}, however, are derived from executions of the system under test. As a result, \textit{faults already present} in the current version are recorded as \textit{expected behavior} rather than exposed. Property-based testing offers \textit{independent semantic oracles} over input domains, but in practice it depends on high-quality properties and provides little guidance for reaching deep program states or satisfying selective preconditions.

We present \tool{} (\underline{PRO}perty-\underline{G}uided \underline{RE}gression \underline{S}earch for \underline{S}emantic Falsification), a testing framework that integrates intent-driven properties directly into coverage-guided, search-based evolutionary test generation, enabling tests to \textit{reach} deep program states and \textit{detect violations of intended behavior}. Concretely, \tool{} proceeds in three steps: (1) it extracts intent-bearing code context and uses a language-model pipeline to generate executable \textsc{jqwik} properties while limiting implementation leakage; (2) it extends EvoSuite's DynaMOSA with one search objective per property and a property-aware fitness function that rewards progress through preconditions and prioritizes falsifying executions; and (3) it binds property parameters, materializes inputs, and uses \textsc{jqwik}-provided generators to connect quantified inputs to evolving test sequences, allowing properties to steer generation toward both coverage and bug-detection goals.

We evaluate \tool{} on 25 large-scale Java systems against regression-test generation, standalone property-based testing, and context ablations. \tool{} detects 328/562 current system version bugs (58\%) while regression-test generation detects none, and satisfies all preconditions for 70/150 hard-to-reach properties versus 18 for standalone \textsc{jqwik}. Ablations show that documentation and caller/callee context are key to generating valid executable properties. Overall, \tool{} preserves structural exploration while exposing current system version faults missed by regression-derived assertions; we release a comprehensive artifact package.

\end{abstract}


\section{Introduction}
\begin{figure*}[t]
    \centering
    \includegraphics[width=\linewidth]{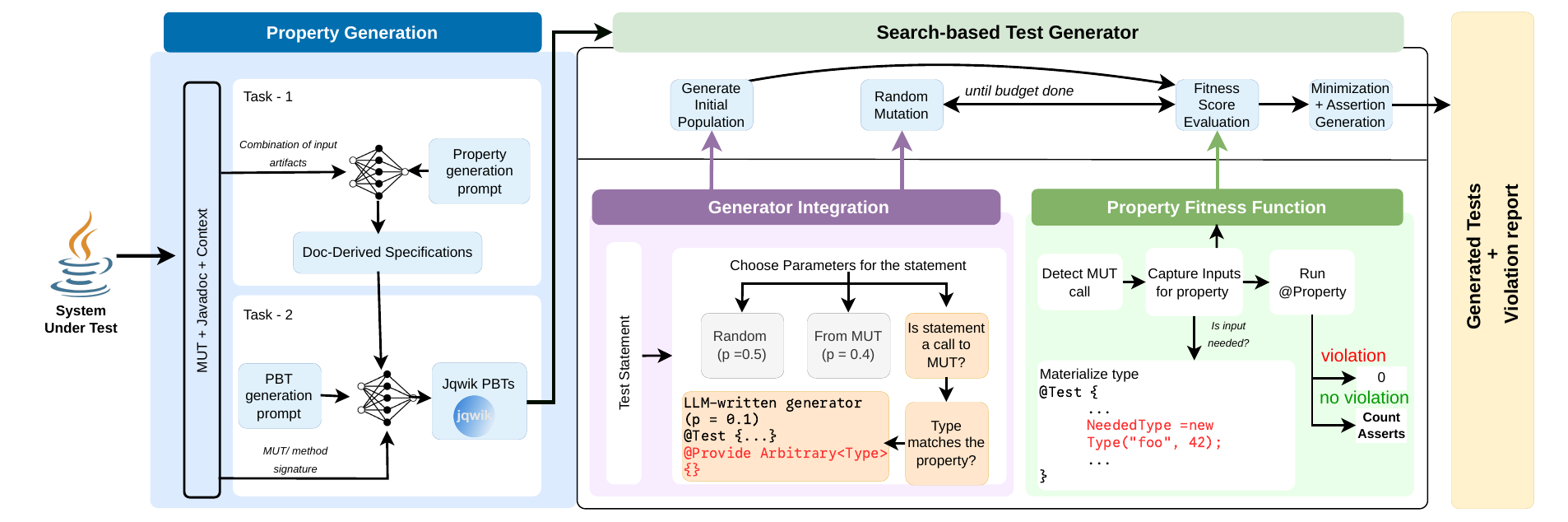}
   \caption{An overview of our \tool{} pipeline.} 
    \label{fig:progress_pipeline}
\end{figure*}

Automated test generation must answer two coupled questions: \emph{where should testing explore?} and \emph{what behavior should hold for those executions?} The first question spans both the input space and the program-structure space: generated inputs and call sequences must drive execution into relevant code regions and program states. \emph{Code coverage} provides one way to guide this exploration; \emph{test oracles} address the second question. Search-based regression testing is highly effective at exploring program structure and producing high-coverage test suites~\cite{Fraser2011EvoSuite}. Its \textit{test oracles}, however, are derived from executions of the current program. Consequently, \textit{buggy behavior} present in the code is encoded as \textit{expected behavior}. This weakness is especially dangerous in AI-assisted software development, where testing the implementation against its own behavior creates a circular process that preserves defects instead of uncovering them.

Property-based testing (PBT) checks general behavioral properties over automatically generated inputs rather than relying on a fixed set of examples~\cite{Claessen2000QuickCheck,GoldsteinEtAl2024}. Because these properties can specify expected behavior independently of the current implementation, PBT has the potential to address the oracle limitation of regression testing. Its effectiveness, however, depends critically on input generation. A major challenge is designing generators that efficiently produce valid, diverse, and fault-revealing inputs~\cite{GoldsteinEtAl2024,LoscherSagonas2017}. If the generated inputs fail to reach relevant program states, even a strong property provides little benefit because it is never evaluated where the implementation violates it.

\begin{minipage}{\linewidth}
\begin{lstlisting}[
  language=Java,
  caption={Regression testing executes the fault but fail to detect it, while standalone PBT struggles to reach the valid input region.},
  label={lst:motivation}
]
@Test
void generatedRegressionTest() {
    Frame f = Parser.parse("@03:abc#7A");
    assertEquals("acb", f.payload()); // observed bug
}

@Property
void validFramesRoundTrip(@ForAll String s) {
    Assume.that(s.startsWith("@"));
    Assume.that(hasValidLength(s));
    Assume.that(hasValidChecksum(s));
    assertThat(encode(Parser.parse(s))).isEqualTo(s);
}
\end{lstlisting}
\end{minipage}

Consider the parser in Listing~\ref{lst:motivation}, whose faulty
implementation swaps two payload characters. Regression-test generation
reaches the fault but records the observed payload \texttt{"acb"} as the
expected result, causing the test to pass. An intent-derived round-trip
property instead exposes the fault without relying on the parser's observed
output as the oracle. However, when standalone property-based testing
generates arbitrary strings, most inputs fail the format assumptions before
reaching the assertion, with no guidance toward valid frames. A custom
generator could solve this problem, but would require manually encoding the
input format and setup logic~\cite{Claessen2000QuickCheck,LinkJqwik}.

The example reveals complementary \textit{limitations}: regression-test generation reaches
the faulty behavior but lacks an independent oracle, whereas property-based
testing supplies the oracle but lacks guidance toward inputs that satisfy its
preconditions. Running them independently preserves both limitations. Prior work,
such as targeted PBT and JQF, guides input generation using heuristic search or
coverage feedback~\cite{LoscherSagonas2017,PadhyeEtAl2019}. However, these
approaches do not fully exploit modern search-based techniques such as
EvoSuite's DynaMOSA, which evolves complete tests, constructs complex call
sequences, and dynamically prioritizes uncovered targets~\cite{Fraser2011EvoSuite,Panichella2018DynaMOSA}.

To address these limitations, we present \tool{} (\underline{Pro}perty-\underline{G}uided \underline{Re}gre\underline{ss}ion Testing), a novel framework that integrates intent-driven properties directly into coverage-guided, search-based test generation. Figure~\ref{fig:progress_pipeline} shows our pipeline: \tool{} first extracts method documentation and code context to generate executable \textsc{jqwik}~\cite{LinkJqwik} properties, then injects these properties into EvoSuite through \textit{property objectives}, \textit{parameter binding}, \textit{generator integration}, and \textit{property-aware fitness}. \tool{} retains DynaMOSA's~\cite{Panichella2018DynaMOSA} structural objectives and adds search objectives for each executable \textsc{jqwik} property. When an evolving test invokes the target method, a property checker captures the relevant values, binds them to the property's quantified parameters, and evaluates the property. The property-aware fitness prioritizes property violations and otherwise rewards candidates for satisfying progressively more preconditions. In Listing~\ref{lst:motivation}, candidates satisfying the prefix condition outrank arbitrary strings, candidates that also satisfy the length condition receive higher fitness, and a valid frame that violates the round-trip property exposes the fault.

To supply these properties, \tool{} mines \emph{intent-driven} specifications from documentation, program context, and related methods identified through the call graph. A two-stage language-model pipeline first generates evidence-grounded natural-language properties and then translates the verified properties into executable \textsc{jqwik} tests. The language model provides semantic objectives, while \tool{} makes them actionable within coverage-guided search.  Section~\ref{sec:methodology} describes our context extraction, property generation and EvoSuite extension in detail.

We evaluate \tool{} on 25 large-scale Java systems through three research questions. First, \tool{} detects 328/562 injected bugs (58\%) missed by regression-derived assertions, while regression-test generation detects none. Second, on 150 hard-to-reach properties, \tool{} satisfies all preconditions for 70 properties, compared with 18 for standalone \textsc{jqwik}. Third, our context ablation shows that documentation and caller/callee context are key to producing valid executable properties. Overall, \tool{} preserves the reachability strengths of search-based testing while adding independent semantic oracles that expose current system version faults.

In summary, this paper makes the following contributions:
\begin{itemize}
    \item We introduce a property-aware extension of DynaMOSA that treats executable properties as first-class search objectives and rewards progress toward falsification.
    
    \item We develop mechanisms for quantified-parameter binding, input materialization, and \textsc{jqwik}-guided value generation.
    
    \item We develop an intent-driven, two-stage pipeline for deriving grounded natural-language and executable properties from method code, documentation, and caller/callee context.
    
    \item We evaluate \tool{} on 25 large-scale Java systems through mutation analysis, comparison with standalone PBT, and context ablations, and release a replication package containing the implementation, prompts, properties, tests, and experimental artifacts.
\end{itemize}

\section{Background}
This section provides background on property-based testing and search-based test generation, the two testing paradigms that \tool{} brings together.

\subsection{Property-Based Testing and \textsc{jqwik}}
\label{sec:intro-pbt}

Property-based testing checks general behavioral properties over automatically generated inputs rather than relying on a fixed set of examples~\cite{Claessen2000QuickCheck,GoldsteinEtAl2024}. A property can be viewed as a falsifiable statement of the form:
\begingroup
\footnotesize
\begin{equation}
\begin{aligned}
& \forall x_1 \in D_1, \dots, x_k \in D_k.\;
\underbrace{P(\mathbf{x})}_{\text{precondition}} \\
& \quad \Longrightarrow\;
\underbrace{Q\bigl(\mathbf{x},
    \operatorname{Exec}(T(\mathbf{x}))\bigr)}
    _{\text{postcondition}},
\end{aligned}
\label{eq:property}
\end{equation}
\endgroup
where $\mathbf{x}=(x_1,\dots,x_k)$ contains values drawn from domains
$D_1,\dots,D_k$. The precondition $P$ identifies admissible inputs,
$T(\mathbf{x})$ denotes the test constructed from them, and
$\operatorname{Exec}(T(\mathbf{x}))$ captures its observable execution.
The postcondition $Q$ specifies the behavior that must hold. Quantified
values need not be direct arguments of the method under test; they can also
construct object state or determine earlier calls in the test.

In this work, we use \textsc{jqwik}, a Java PBT engine built on the JUnit Platform%
~\cite{LinkJqwik}. A method annotated with \texttt{@Property} defines a
property, its parameters, annotated with \texttt{@ForAll}, define its domain. Each
parameter receives values from an \texttt{Arbitrary<T>}. \textsc{jqwik} provides
default arbitraries for common Java types and supports domain-specific
arbitraries through \texttt{@Provide} methods. Within a property,
\texttt{Assume.that(\dots)} expresses the precondition $P$, while assertions express the postcondition $Q$.
Inputs violating an assumption are discarded; when an assertion fails,
\textsc{jqwik} attempts to shrink the input into a simpler counterexample.

PBT therefore depends critically on its generators. In Listing~\ref{lst:motivation}, arbitrary strings rarely satisfy the prefix,
length, and checksum assumptions, so most executions are discarded before
reaching the round-trip assertion. A custom arbitrary could encode valid
frames, but constructing valid, diverse, and fault-revealing generators
requires manual effort and domain knowledge~\cite{GoldsteinEtAl2024}.
Consequently, even a strong property provides little benefit when generation
does not reach the program states in which it can be falsified.

We selected \textsc{jqwik} because its API exposes the components required by our
integration: quantified parameters identify values to bind, arbitraries
provide generation knowledge, assumptions expose preconditions, and
assertions define behavioral objectives. Also, to evaluate our technique, we need tools that support the Java language and the JUnit testing framework, both of which are supported by \textsc{jqwik}. This allows \tool{} to reuse values
constructed by evolving EvoSuite tests, invoke \textsc{jqwik} generators if domain-specific knowledge is needed, and use preconditions and property violations
to guide the search.

\subsection{EvoSuite's DynaMOSA Search}
\label{sec:intro-evosuite}

\begin{figure}[h]
\centering
\resizebox{\columnwidth}{!}{%
\begin{tikzpicture}[node distance=3mm and 4mm]
  \node[evobox] (gen) {Generate initial\\population};
  \node[evobox, right=of gen] (eval)
        {Execute tests, score active fitness functions};
  \node[evobox, right=of eval] (select)
        {DynaMOSA ranking and selection)};
  \node[evobox, right=of select] (vary)
        {Crossover\\and mutation};

  \node[evobox, below=5mm of select] (post)
        {Minimize tests and\\generate assertions};
  \node[evobox, right=of post] (out) {JUnit\\suite};

  \draw[evoflow] (gen) -- (eval);
  \draw[evoflow] (eval) -- (select);
  \draw[evoflow] (select) -- (vary);

  \draw[evoflow]
        (vary.north) to[out=90,in=90,looseness=0.8]
        node[above,font=\scriptsize] {next generation}
        (eval.north);

  \draw[evoflow] (select.south) --
        node[right,font=\scriptsize] {budget exhausted}
        (post.north);
  \draw[evoflow] (post) -- (out);
\end{tikzpicture}%
}
\caption{Overview of DynaMOSA search and post-processing.}
\label{fig:compact-evosuite-dynamosa}
\end{figure}
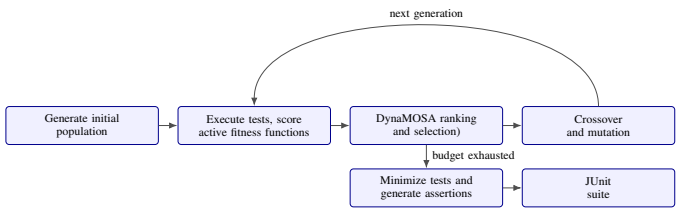

EvoSuite~\cite{Fraser2011EvoSuite} is a search-based unit-test generator for
Java. Given a class under test, EvoSuite evolves coverage-optimized test cases
and produces a minimized JUnit suite with regression assertions. Under
DynaMOSA, each candidate test is represented internally as a variable-length
\emph{chromosome} containing primitive assignments, constructors, field
accesses, and method calls. Figure~\ref{fig:compact-evosuite-dynamosa}
summarizes the corresponding evolutionary loop.

As an example, Listing~\ref{lst:intstack-evolution} shows a test EvoSuite evolved for the \texttt{IntStack} class. The
comments (added for illustrative purposes) indicate which phase introduces each part of the test.
\begin{minipage}{\linewidth}
\begin{lstlisting}[
  caption={One possible evolution of \texttt{test4}.},
  label={lst:intstack-evolution}
]
@Test(timeout = 4000)
public void test4() throws Throwable {
    IntStack iStk0 = new IntStack();   // generation
    iStk0.pushTwo(0, 0);               // generation
    int int0 = iStk0.pop();             // mutation

    assertEquals(1, iStk0.size()); // observed state
    assertEquals(0, int0);        // observed output
}
\end{lstlisting}
\end{minipage}

The loop proceeds as follows.

\begin{enumerate}[topsep=2pt,itemsep=2pt,leftmargin=*]
  \item \textbf{Generation.}
  EvoSuite begins with a population of random tests, consisting of a sequence of random method or constructor calls. When a selected constructor
  or method requires an argument of type~$T$, EvoSuite reuses a compatible value
  already in the test or recursively creates one from the test cluster. Thus, a
  single method call can introduce several supporting statements. In
  Listing~\ref{lst:intstack-evolution}, the \texttt{pushTwo} call recursively introduces the \texttt{IntStack} constructor.

  \item \textbf{Fitness evaluation.}
  Each candidate test is executed, and runtime observation is used to evaluate coverage (fitness) along multiple dimensions, such as branch, exception, and line. DynaMOSA~\cite{Panichella2018DynaMOSA} dynamically activates branch-related fitness functions only when their parent in the control-flow graph is covered.

  \item \textbf{DynaMOSA ranking.}
  DynaMOSA~\cite{Panichella2018DynaMOSA} ranks individual tests via a preference criterion which ensures each active fitness function is optimized simultaneously, and difficult-to-reach fitness functions are prioritized first. Pareto fronts are constructed by successively selecting all non-dominated tests (i.e. there is not another test with strictly greater fitness in every function), and crowding distance is used as a tie-breaker in the final front. These tests are passed to the next step.

  \item \textbf{Crossover and mutation.}
  Selected parent tests produce new candidates through crossover and mutation.
  Crossover combines portions of two tests, while mutation inserts, deletes, or
  changes statements and primitive values. In Listing~\ref{lst:intstack-evolution}, an insertion
  or crossover can add the call to \texttt{pop}, which will be preferred during the next selection step due to exercising more behavior than before. The resulting offspring form
  the next generation and are evaluated again.

  \item \textbf{Post-processing.}
  When the search budget is exhausted, EvoSuite removes tests and statements
  that do not contribute coverage. It then adds assertions over observed return
  values and object states, producing the final regression suite. In Listing~\ref{lst:intstack-evolution},
  the two assertions are generated here.
\end{enumerate}

Importantly, these final assertions record behavior observed from the current implementation, which could include unintended behavior. This motivates \tool{}'s extension: DynaMOSA provides strong structural reachability, but regression assertions alone do not provide independent semantic oracles.

\section{Implementation}
\label{sec:methodology}

\subsection{EvoSuite Extension}
\label{sec:evosuite}

\tool{} extends EvoSuite in \textit{three} ways:
(1)~a property fitness function;
(2)~materialization of property inputs so they can be evolved; and
(3)~a hook to use
\texttt{@Provide} generators. 

Below, we discuss each step. We use the following \textsc{jqwik} property as our running example throughout this section:
\begin{lstlisting}[caption={Running property for \texttt{IntStack\#pop}.},label={lst:propexample}]
@net.jqwik.api.Property
@TargetMethod("com.evopbt.demo.IntStack#pop#()I")
void popAfterPushTwo(@ForAll int first, @ForAll int second) {
    Assume.that(second != Integer.MIN_VALUE);      // precondition P
    IntStack stack = new IntStack();
    stack.pushTwo(first, second);
    assertThat(stack.pop()).isEqualTo(second);     // postcondition Q
}
\end{lstlisting}

\subsubsection{Property fitness}
\label{sec:fitness-hl}

Each \texttt{@Property} is linked to the MUT it checks via a \texttt{@TargetMethod} annotation. Whenever a generated test calls that MUT, the property is checked, resulting in a fitness score~$s$ given to
DynaMOSA (lower is better):

\begingroup
\footnotesize
\begin{equation}
s \;=\;
\begin{cases}
0 & \text{assertion failed (violation detected)}\\[2pt]
T - p + 1 &
  \begin{minipage}[t]{0.62\columnwidth}\raggedright\footnotesize
  no assertion failed, and $p$ of the $T$ preconditions (\texttt{Assume.that()}) passed; if $p \neq T$, the $(p{+}1)$-th precondition failed.
  \end{minipage}\\[2pt]
T + 2 & %i updated the code but didn't update this section
  \begin{minipage}[t]{0.62\columnwidth}\raggedright\footnotesize
  unexpected assumption threw (buggy property)
  \end{minipage}
\end{cases}
\label{eq:score}
\end{equation}
\endgroup

where $T$ is the number of \texttt{Assume.that} calls in the body.

\textbf{Example:} The property has $T=1$.
If the only
precondition fails, $s=2$. If it passes, $s=1$. 
If the search finds an input that makes the assertion fail, $s=0$. This score gives a helpful slope when each precondition can be reached, but the combination of preconditions is hard to reach.

\subsubsection{Input materialization}
\label{sec:param-binding}
Listing~\ref{lst:propexample} quantifies over two \texttt{int}s. During mutation, these parameters are created (if they don't already exist) before the call to the MUT. During test replay, these additional parameters are captured and fed into the property.

\textbf{Example:} Two \texttt{int} statements are needed to check \texttt{popAfterPushTwo}, and will be inserted into
\texttt{test4} before the \texttt{pop()} call. When \texttt{pop()} runs, \texttt{PropertyChecker} uses
these \texttt{int} values, passing them as to \texttt{first} and
\texttt{second} to the property body:
\begin{lstlisting}[caption={\texttt{test4}: needed \texttt{int}s materialized; \texttt{pushTwo} args harvested as \texttt{first}/\texttt{second} at \texttt{pop()}.},label={lst:binding-example}]
int int1 = 0;
int int2 = 0;
IntStack intStack0 = new IntStack();
intStack0.pushTwo(int1, int2);
int int0 = intStack0.pop();
\end{lstlisting}
\subsubsection{Generator hook}
\label{sec:property-generators}
Sometimes, a constructor or method needs specific input to trigger behavior that EvoSuite's search-based machinery is not able to find. For any method with an associated property and generator, EvoSuite use the \textsc{jqwik} \texttt{@Provide} with a configurable probability to create input values.

\textbf{Example:} for \texttt{pushTwo(int,\,int)}, the \texttt{smallInts} arbitrary can be used:
\begin{lstlisting}[caption={Property with \texttt{@Provide} generators for \texttt{pushTwo}.},label={lst:generator-example}]
@Provide
Arbitrary<Integer> smallInts() {
    return Arbitraries.integers().between(0, 10);
}
@Property
@TargetMethod("com.evopbt.demo.IntStack#pushTwo#(II)V")
void pushTwoPreservesTop(@ForAll("smallInts") int a, @ForAll("smallInts") int b) {
    IntStack stack = new IntStack();
    stack.pushTwo(a, b);
    assertThat(stack.pop()).isEqualTo(b);
}
\end{lstlisting}
\noindent This biases EvoSuite towards values from the generator such as \texttt{pushTwo(3,\,7)}.

%It is a bias, since it is a probability to generate these values

\subsection{Context construction.}
\label{sec:ctx}
As part of \tool{}'s property-generation implementation, we construct a
context bundle for each method under test. The bundle contains the focal
method code or signature, method documentation, enclosing-class documentation,
and call-graph context: the fully qualified method signature, code and documentation of resolved callers and
callees. We implement this with a Java extractor that parses source code and
structured Javadocs with \textsc{Spoon}~\cite{spoon}, constructs a static call graph with
\textsc{SootUp}~\cite{sootup}, and reconciles call-graph edges with source-level method
identities when possible. Each artifact is emitted with an origin label, such as
\texttt{method\_code}, \texttt{method\_doc}, \texttt{class\_doc (cd)},
\texttt{caller\_code (cac)}, \texttt{caller\_doc (cad)}, \texttt{callee\_code (cec)}, and
\texttt{callee\_doc (ced)}. These labels allow \tool{} to assemble different prompt
configurations for context ablations and to keep documentation-derived evidence
separate from implementation-derived evidence.

The selected context bundle is then passed to the language-model pipeline:
the first stage derives natural-language behavioral properties grounded in the
supplied artifacts, and the second translates accepted properties into
executable \textsc{jqwik} properties as shown in Figure~\ref{fig:progress_pipeline}.

\subsection{Property Generation}
\label{sec:propgen}

\tool{} uses a language-model pipeline to turn program intent into executable
\textsc{jqwik} properties. For each MUT, the pipeline receives a context bundle
constructed from some combination of the method body or signature, the method
Javadoc, enclosing-class documentation, and the code and documentation of
resolved callers and callees. Each artifact is tagged with its origin, enabling
context ablations and allowing generated properties to be traced back to the
evidence that supported them.

We use Opus 4.8 within Claude Code as the LLM interface for property generation. We keep the LLM backend fixed across all experiments because our objective is to evaluate the value of property-guided search, not to rank models for property-based test synthesis. Holding the model fixed makes the main experimental variables those central to \tool{}: context selection, prompt structure, and the integration of generated properties into EvoSuite's search.

\paragraph{Two-stage prompting}

Our pipeline separates \emph{specification mining} from
\emph{code synthesis}. In Stage~1, the LLM receives the selected context bundle
and produces a numbered list of natural-language behavioral claims grounded
strictly in the supplied artifacts. Compound claims are split into individual properties, and each claim carries a source label indicating which artifacts support it: the method body alone, the body together with its docstring, or the full surrounding context.

Stage~2 receives the accepted Stage~1 claims, together with inputs depending on the input configuration, and translates each claim into
one or more executable \textsc{jqwik} \texttt{@Property} methods. 

We vary the artifact subset to measure which forms of context help produce
useful property objectives, discussed in Section~\ref{sec:RQ3}.

\paragraph{Prompt constraints}

The pipeline enforces constraints needed by \tool{}'s search-based execution
model. First, every generated \texttt{@Property} must contain at least one
\texttt{Assume.that(\ldots)} precondition before assertions, so the property-aware
fitness function can reward partial progress through preconditions instead of
treating unsatisfied assumptions as uninformative failures. Second, each
\texttt{@Property} must encode exactly one behavioral claim, allowing each
property to correspond to a distinct DynaMOSA search objective. Third, custom
\texttt{@Provide} generators are preferred over unconstrained \texttt{@ForAll}
inputs when domain constraints, valid ranges, or structural invariants can
reduce discarded inputs.

\section{Evaluation}
\label{sec:evaluation}

We evaluate \tool{} along three dimensions: whether intent-driven properties
effectively detect current system version faults, whether integration with search-based test generation provides an advantage over standalone property-based testing,
and how context selection affects the usefulness of generated properties.

In this section, we answer the following research questions:

\textbf{RQ1.} \emph{Can intent-driven properties generated by \tool{} detect faults in the current program version that regression-test generation misses?}

\textbf{RQ2.} \emph{Does property-guided search exercise and falsify generated properties more effectively than standalone property-based testing?}

\textbf{RQ3.} \emph{How does the choice of property-generation context affect property validity and fault-detection effectiveness?}

\subsection{RQ1: Detecting Bugs in the Current Version}
\label{sec:rq1}

RQ1 evaluates the central oracle claim of \tool{}: regression-test generation can execute faulty behavior but encode it as expected behavior, while \tool{} can use intent-derived properties as test oracles independent of the current implementation. We therefore ask how effectively generated properties can detect current system version faults that regression-derived assertions miss.

\paragraph{Systems Under Test.}
To evaluate \tool{} on large-scale real-world systems, we use the OE25 systems~\cite{togaeval,togll,konstantinou2024llms,TasnimDwyerHossain2026TOGBench}, a corpus widely used in testing research, consisting of 25 Java systems: 8 from the EvoSuite benchmark and 17 Apache Commons packages. We select MUT--Javadoc pairs through a three-stage filtering pipeline. Deterministic filters retain methods and Javadocs with sufficient length, control-flow complexity, and documentation quality, yielding 538 candidates.

The same fixed backend (Opus~4.8, \S\ref{sec:propgen}) then scores candidates for documentation alignment~(R1), input--output clarity~(R2), and PBT property derivability~(R3), weighted 0.30, 0.30, and 0.40, respectively, while excluding methods with external I/O, reflection or dynamic class loading, no observable test oracle, or nondeterminism. Candidates with score $\geq 0.60$ and no exclusion flags yield 279 samples. Since the scorer shares the generation backend, the filter may favor the generator's own conventions; the executable-validation step and the two-author manual review bound this risk. Weights and threshold were fixed a priori and not adjusted after observing evaluation outcomes.

Finally, two authors manually review the remaining candidates to remove trivial, vague, network-dependent, or implicitly nondeterministic cases, producing 240 high-quality, PBT-suitable Java method--Javadoc pairs.

\begin{table}[h]
\centering
\small
\caption{MUT-level execution outcomes}
\setlength{\tabcolsep}{6pt}
\begin{tabular}{@{}lrr@{}}
\toprule
\textbf{MUT-level outcome} & \textbf{Count} & \textbf{\%} \\
\midrule
Failed runs                         & 50  & 21   \\
\quad Compilation failure            & 15  & 6   \\
\quad EvoSuite crash                 & 1   & 0   \\
\quad No valid properties generated & 34 & 14 \\
MUTs with no property violation      & 147 & 61  \\
MUTs with $\geq 1$ property violation & 43  & 18 \\
\midrule
\textbf{Total}                       & 240 & 100 \\
\bottomrule
\end{tabular}
\label{tab:rq1_mut_summary}
\end{table}

\begin{table}[h]
\centering
\small
\caption{Property-level validation on the unmutated baseline.}
\setlength{\tabcolsep}{6pt}
\begin{tabular}{@{}lrr@{}}
\toprule
\textbf{Property-level outcome} & \textbf{Count} & \textbf{\%} \\
\midrule
Baseline-valid properties       & 816 & 93  \\
Baseline-violating properties   & 64  & 7   \\
\midrule
\textbf{Total}                  & 880 & 100 \\
\bottomrule
\end{tabular}

\label{tab:rq1_property_summary}
\end{table}

\paragraph{Experimental Setup.}
We follow a three-step pipeline. First, we inject source-level mutants using PITMuS and PIT mutation testing~\cite{Hossain2026PITMuS,coles2016pit}. For each mutant, PITMuS produces a mutated Java source file that we compile and treat as the current system version. We record the injected bug type, mutant counts, and system metadata. Second, we generate executable \textsc{jqwik} properties using the pipeline in Section~\ref{sec:propgen}, with a fixed Claude Code/Opus 4.8 backend: Stage~1 derives evidence-grounded natural-language properties, and Stage~2 translates those properties into executable \textsc{jqwik} properties. We also record the metadata needed to integrate each property with the SUT. Third, \tool{} runs these properties on the mutated code to detect property violations. Before mutation evaluation, we run each property on the unmutated baseline and discard any property that already fails there: such a property cannot distinguish a fault introduced by the mutant from an invalid oracle. We call the remaining properties \emph{baseline-valid}. A mutant is killed when at least one baseline-valid property passes on the original version but fails on the mutant. Table~\ref{tab:rq1_mut_summary} and ~\ref{tab:rq1_property_summary} report execution statistics, and Table~\ref{tab:rq1_mutation} reports property violations for the injected bugs.

\begin{table}[h]
\centering
\small
\caption{Mutation-level detection results. Rows by mutation operator report \tool{} results; the final rows compare \tool{} with EvoSuite overall.}
\setlength{\tabcolsep}{5pt}
\begin{tabular}{@{}lrrr@{}}
\toprule
\textbf{Mutation} & \textbf{Mutants} & \textbf{Killed} & \textbf{Kill (\%)} \\
\midrule
Negated cond.          & 236 & 168 & 71  \\
Conditional bound.     & 80  & 16  & 20  \\
Math                   & 66  & 20  & 30  \\
Null return            & 60  & 29  & 48  \\
Other return value     & 55  & 47  & 85  \\
Void method call       & 36  & 23  & 64  \\
Boolean true ret.      & 13  & 12  & 92  \\
Boolean false ret.     & 7   & 5   & 71  \\
Other                  & 5   & 5   & 100 \\
Increment/decre.       & 4   & 3   & 75  \\
\midrule
\textbf{\tool{}}       & 562 & 328 & 58  \\
\textbf{Regression Test}      & 562 & 0   & 0   \\
\bottomrule
\end{tabular}

\label{tab:rq1_mutation}
\end{table}

\subsection{Results}

Table~\ref{tab:rq1_mut_summary} reports MUT-level baseline validation before mutation evaluation. Of the 240 MUTs, 50 are non-evaluable because no valid properties are generated, a \texttt{javac} compilation error occurs, or EvoSuite throws an error. Among the remaining 190 evaluable MUTs, 147 (77\%) have no property violation on the unmutated baseline, while 43 (23\%) have at least one violation-triggering property that was discarded.

Table~\ref{tab:rq1_property_summary} shows the properties that are generated for the 147 MUTs. A single MUT can have multiple properties. Of 880 generated properties, 816 pass on the unmutated baseline and are retained for mutation evaluation, while 64 fail on the baseline and are treated as false positives. This 93\% baseline-valid rate indicates that the property-generation pipeline usually produces executable oracles that do not reject the original version.

Table~\ref{tab:rq1_mutation} reports bug-detection results for the injected source-level faults, grouped by mutation operator.  For each retained MUT, PITMuS/PIT can generate multiple mutants by applying different mutation operators at different mutation sites in the source code. For example, a single statement such as \texttt{if (x >= 0) return x + y;} can yield separate mutants: \texttt{if (x < 0) return x + y;} for a negated conditional, \texttt{if (x > 0) return x + y;} for a conditional-boundary change, and \texttt{if (x >= 0) return x - y;} for a math-operator change. Across the 147 retained MUTs, this process produces 562 evaluated source-level mutants, about 3.8 mutants per MUT on average. \tool{} detects 328 of 562 injected bugs, for an overall detection rate of 58\%. These results show that intent-derived properties can serve as independent oracles: when the mutated program violates the intended behavior expressed by a property, \tool{} can expose the bug.

Detection is strongest for bugs that directly affect observable behavior. For example, \tool{} detects 47 of 55 \emph{other return value} bugs and 12 of 13 \emph{boolean true return} bugs. It also detects 168 of 236 \emph{negated conditional} bugs, the largest absolute number of detections. Detection is lower for \emph{conditional boundary} and \emph{math} bugs, with rates of 20\% and 30\%, respectively, since these often require properties that capture narrow boundary conditions or precise numerical relationships.

In contrast, regression-test generated by EvoSuite detects 0 of 562 injected bugs. This does not mean the generated regression tests cannot execute the faulty code; rather, their assertions are derived from the mutated version itself. As a result, the bug is recorded as expected behavior instead of being exposed. 
The comparison therefore highlights the key value of \tool{}: generated properties give search-based testing an independent semantic oracle that can reveal current system version bugs missed by regression-derived assertions.

\begin{rqanswerbox}{Answer to RQ1}
\tool{} detects current system version bugs that regression-test generation misses.
\tool{} detects 328 of 562
source-level injected bugs (58\%), while EvoSuite detects none in this setting.
These results show that intent-derived properties provide independent semantic
oracles for bugs that regression assertions can preserve.
\end{rqanswerbox}

\subsection{RQ2: Reaching Hard Property Preconditions}
\label{sec:rq2}

RQ2 evaluates the reachability claim of \tool{}. Regular property-based testing can provide semantic oracles, but those oracles are useful only when generated inputs satisfy the properties' preconditions. \tool{} \textit{extends} EvoSuite's DynaMOSA search and \textit{integrates} properties into the search process, allowing object construction, call sequences, and generated values to evolve toward precondition satisfaction. We therefore ask whether \tool{} reaches hard-to-reach property preconditions more effectively than standalone \textsc{jqwik}.

\paragraph{Experimental Setup.}
RQ2 uses OE25 systems~\cite{togaeval,togll,konstantinou2024llms,TasnimDwyerHossain2026TOGBench}, but filters for properties whose preconditions are hard to satisfy. We retain non-primitive types with a public constructor or factory method (518 candidate types, with 5232 possible methods) and randomly select 500 methods that take such a type as a parameter. We generate properties for them with 381 methods yielding properties, and keep samples where \textsc{jqwik} needs more than 10 attempts to satisfy the first precondition. This yields exactly 150 samples with hard-to-reach preconditions (Each sample is a property-MUT pair). We compare standalone \textsc{jqwik} with \tool{} under equivalent budgets (3 minute wall clock time on the same properties: \textsc{jqwik} uses default or generated arbitraries, while \tool{} integrates the properties into its property-guided DynaMOSA search. We instrument preconditions and assertions to measure first-precondition satisfaction, all-precondition satisfaction, assertion reachability, and time; Tables~\ref{tab:rq2_assertion_reach}, \ref{tab:rq2_precondition_pass_rate}, and \ref{tab:rq2_assertion_timing} report these results.

\begin{table}[h]
\centering
\caption{Precondition reachability over 150 hard-property MUT samples.}
\label{tab:rq2_assertion_reach}
\small
\setlength{\tabcolsep}{6pt}
\begin{tabular}{@{}lrr@{}}
\toprule
\textbf{Metric} & \textbf{\textsc{jqwik}} & \textbf{\tool{}} \\
\midrule
Rows with first precondition satisfied  & 19 & 70 \\
Rows with all preconditions satisfied   & 18 & 70 \\
\bottomrule
\end{tabular}
\end{table}

\begin{table}[h]
\centering
\caption{Precondition pass rates when an \texttt{Assume.that} line is reached, computed as \texttt{pass\_count / reach\_count} and aggregated over all MUTs.}
\label{tab:rq2_precondition_pass_rate}
\small
\setlength{\tabcolsep}{6pt}
\begin{tabular}{@{}lrr@{}}
\toprule
\textbf{Metric} & \textbf{\textsc{jqwik}} & \textbf{\tool{}} \\
\midrule
First precondition pass rate (\%) & 3.7 & 33.5 \\
All preconditions pass rate (\%)  & 3.6 & 34.0 \\
\bottomrule
\end{tabular}

\end{table}

\begin{table}[h]
\centering
\caption{Median time to assertion passes. \tool{} timing is measured from the start of initial population generation.}
\label{tab:rq2_assertion_timing}
\small
\setlength{\tabcolsep}{6pt}
\begin{tabular}{@{}lrr@{}}
\toprule
\textbf{Metric} & \textbf{\textsc{jqwik}} & \textbf{\tool{}} \\
\midrule
Median time to first assertion pass (s) & 0.014 & 1.541 \\
Median time to all assertions pass (s)  & 0.105 & 1.615 \\
\bottomrule
\end{tabular}
\end{table}

\paragraph{Results.}
Table~\ref{tab:rq2_assertion_reach} shows that \tool{} reaches hard property preconditions substantially more often than standalone \textsc{jqwik}. Out of 150 hard-property MUTs, \tool{} satisfies the first precondition for 70 MUTs, compared with 19 for \textsc{jqwik}. \tool{} also satisfies all preconditions for 70 MUTs, compared with 18 for \textsc{jqwik}. Thus, the EvoSuite-based search in \tool{} makes many properties applicable that standalone random generation rarely reaches.

Table~\ref{tab:rq2_precondition_pass_rate} shows the same trend at the invocation level. When an assumption is reached, \tool{} satisfies the first precondition in 33.5\% of cases and all preconditions in 34\% of cases, compared with 3.7\% and 3.6\% for \textsc{jqwik}. 
This highlights the key contribution of \tool{}: search-based testing can use preconditions as signals to explore more meaningful program states, and mutate on tests that already trigger meaningful behavior. The set difference
$\textsc{jqwik}_{\mathit{reached}} \setminus \tool{}_{\mathit{reached}}$
contains only five non-comparable \tool{} executions: four underlying EvoSuite
crashes and one run that did not complete input generation. After excluding
these failures, every property reached by standalone \textsc{jqwik} is also
reached by \tool{}.

Table~\ref{tab:rq2_assertion_timing} shows the expected trade-off. Standalone \textsc{jqwik} is faster when its generators already produce admissible inputs, but \tool{} reaches substantially more hard preconditions. Thus, \tool{} trades additional search time for reachability: it spends more effort on instrumentation and constructing call sequences but that effort allows exploration of states that regular PBT often fails to generate.

\paragraph{Qualitative Examples.}
The following examples illustrate when \tool{}'s reachability advantage appears, when both approaches succeed, and when both still fail.

\textit{Example 1: Both pass.}
Property index~30 tests \texttt{ExecutionVisitor.visitASTORE}. The method pops a reference from the operand stack and stores it in a local variable slot. The generated property builds a \texttt{Frame}, pushes \texttt{Type.STRING}, calls \texttt{visitASTORE}, and checks that the stack shrinks by one and the local slot holds the pushed type. The property uses an \texttt{Assume.that} gate to restrict the index to the valid range $[0,16)$. The MUT and generated property are shown in Listings~\ref{lst:rq2-prop30-mut} and~\ref{lst:rq2-prop30-property}.

Both engines succeed for this property because the input constraint is relatively simple. Standalone \textsc{jqwik} passes after its generator draws a valid \texttt{ASTORE} index. \tool{} also finds valid indices at scale, producing 74{,}828 assumption passes and 74{,}828 assertion passes out of 95{,}603 property invocations that reach the assumption.

\begin{lstlisting}[
language=Java,
caption={MUT (property index~30): \texttt{ExecutionVisitor.visitASTORE}.},
label={lst:rq2-prop30-mut}
]
public void visitASTORE(final ASTORE o) {
    locals().set(o.getIndex(), stack().pop());
}
\end{lstlisting}

\begin{lstlisting}[
language=Java,
caption={Generated property (property index~30): \texttt{visitASTORE\_popsStackValueIntoLocalSlot}.},
label={lst:rq2-prop30-property}
]
@Property
void visitASTORE_popsStackValueIntoLocalSlot(
        @ForAll("naiveAstore") ASTORE astore) {
    int idx = astore.getIndex();
    Assume.that(idx >= 0 && idx < MAX_LOCALS);

    Frame frame = new Frame(MAX_LOCALS, MAX_STACK);
    frame.getStack().push(Type.STRING);

    ExecutionVisitor ev = new ExecutionVisitor();
    ev.setFrame(frame);

    int stackSizeBefore = frame.getStack().size();
    ev.visitASTORE(astore);

    assertThat(frame.getStack().size())
        .isEqualTo(stackSizeBefore - 1);
    assertThat(frame.getLocals().get(idx))
        .isEqualTo(Type.STRING);
}

@Provide
Arbitrary<ASTORE> naiveAstore() {
    return Arbitraries.integers().map(ASTORE::new);
}
\end{lstlisting}

\textit{Example 2: \tool{} outperforms \textsc{jqwik}.}
Property index~4 tests \texttt{Cookie.toString(JSONObject)}, which serializes a cookie JSON object to a \texttt{name=value} string and parses it back. The property requires the input \texttt{JSONObject} to contain a non-blank \texttt{"name"} key before checking that \texttt{Cookie.toJSONObject(Cookie.toString(jo))} preserves the trimmed name. The MUT and generated property are shown in Listings~\ref{lst:rq2-prop4-mut} and~\ref{lst:rq2-prop4-property}.

Standalone \textsc{jqwik} reaches the assumption on every attempt but never passes it: its generator inserts a random alphabetic key--value pair, so \texttt{jo.has("name")} is almost never true. As a result, \textsc{jqwik} records 1{,}000 assumption reaches, 0 passes, and 1{,}000 skips. A hand-written generator that always sets \texttt{jo.put("name", ...)} would satisfy the gate immediately, but this requires domain knowledge in the general case. In contrast, \tool{} synthesizes JSON objects with a \texttt{"name"} field and satisfies the round-trip property, producing 5{,}100 assumption passes and 5{,}084 assertion passes.

\begin{lstlisting}[
language=Java,
caption={MUT (property index~4): \texttt{Cookie.toString} (excerpt).},
label={lst:rq2-prop4-mut}
]
public static String toString(JSONObject jo) throws JSONException {
    // ... extract trimmed "name" and "value" keys ...
    if (name == null || "".equals(name.trim())) {
        throw new JSONException("Cookie does not have a name");
    }
    sb.append(escape(name));
    sb.append("=");
    sb.append(escape((String) value));
    // ... append remaining cookie attributes ...
    return sb.toString();
}
\end{lstlisting}

\begin{lstlisting}[
language=Java,
caption={Generated property (property index~4): \texttt{roundTripPreservesName}.},
label={lst:rq2-prop4-property}
]
@Property
void roundTripPreservesName(
        @ForAll("naiveJSONObject") JSONObject jo) {
    Assume.that(jo.has("name"));
    Assume.that(jo.opt("name") instanceof String);
    Assume.that(!((String) jo.opt("name")).trim().isEmpty());

    String cookieStr = Cookie.toString(jo);
    JSONObject parsed = Cookie.toJSONObject(cookieStr);

    assertThat(parsed.getString("name"))
        .isEqualTo(jo.getString("name").trim());
}

@Provide
Arbitrary<JSONObject> naiveJSONObject() {
    return Combinators.combine(
        Arbitraries.strings().alpha().ofMaxLength(8),
        Arbitraries.strings().alpha().ofMaxLength(20)
    ).as((k, v) -> {
        JSONObject jo = new JSONObject();
        jo.put(k, v);
        return jo;
    });
}
\end{lstlisting}

\textit{Example 3: Both fail.}
Property index~0 tests \texttt{CDL.toJSONArray(names, csvString)}: parse a CSV string into a \texttt{JSONArray} of \texttt{JSONObject}s using \texttt{names} as column headers. The property gates on a non-empty \texttt{names} array and a \texttt{csvString} for which the method returns a non-null result (at least one valid data row); it then checks that every key in every output row appears in \texttt{names}. The MUT and generated property are shown in Listings~\ref{lst:rq2-prop0-mut} and~\ref{lst:rq2-prop0-property}.

Neither engine passes all preconditions.
Standalone \textsc{jqwik}'s blind \texttt{@Provide} parses random strings into \texttt{JSONArray}s and often yields empty arrays, so the first gate rarely holds; random \texttt{String} inputs almost never form valid CSV with a matching row.
\tool{} tries more inputs but still never satisfies both gates together---the precondition encodes a brittle format constraint with no constructive API to build admissible \texttt{(names, csvString)} pairs.
This case illustrates a remaining limitation: when a generated precondition is too sharp or depends on a precise input format, additional generator knowledge may still be needed.

\begin{lstlisting}[
language=Java,
caption={MUT (property index~0): \texttt{CDL.toJSONArray} (excerpt).},
label={lst:rq2-prop0-mut}
]
public static JSONArray toJSONArray(JSONArray names, JSONTokener x)
        throws JSONException {
    if (names == null || names.length() == 0) {
        return null;
    }
    JSONArray ja = new JSONArray();
    // ... parse rows until rowToJSONObject returns null ...
    if (ja.length() == 0) {
        return null;
    }
    return ja;
}
\end{lstlisting}

\begin{lstlisting}[
language=Java,
caption={Generated property (property index~0): \texttt{toJSONArrayKeysAreSubsetOfNames}.},
label={lst:rq2-prop0-property}
]
@Property
void toJSONArrayKeysAreSubsetOfNames(
        @ForAll("naiveJSONArray") JSONArray names,
        @ForAll String csvString) {
    Assume.that(names.length() > 0);
    JSONArray result = CDL.toJSONArray(names, csvString);
    Assume.that(result != null);
    // every key in every result row must appear in names
    // ...
}
@Provide
Arbitrary<JSONArray> naiveJSONArray() {
    return Arbitraries.strings().map(s -> {
        try { return new JSONArray(s); }
        catch (Exception e) { return new JSONArray(); }
    });
}
\end{lstlisting}

\begin{rqanswerbox}{Answer to RQ2}
\tool{} reaches hard property preconditions much more effectively than standalone \textsc{jqwik}. Across 150 hard-property rows, \tool{} satisfies all preconditions for 70 MUTs, compared with 18 for \textsc{jqwik}; at the invocation level, \tool{} satisfies all preconditions in 34\% of reached cases, compared with 3.6\% for \textsc{jqwik}. These results show that \tool{}'s property-guided DynaMOSA extension turns preconditions into search guidance, enabling generated semantic oracles to execute in states that standalone PBT often fails to generate.
\end{rqanswerbox}

\begin{figure*}[h]
\centering
\includegraphics[width=0.9\textwidth]{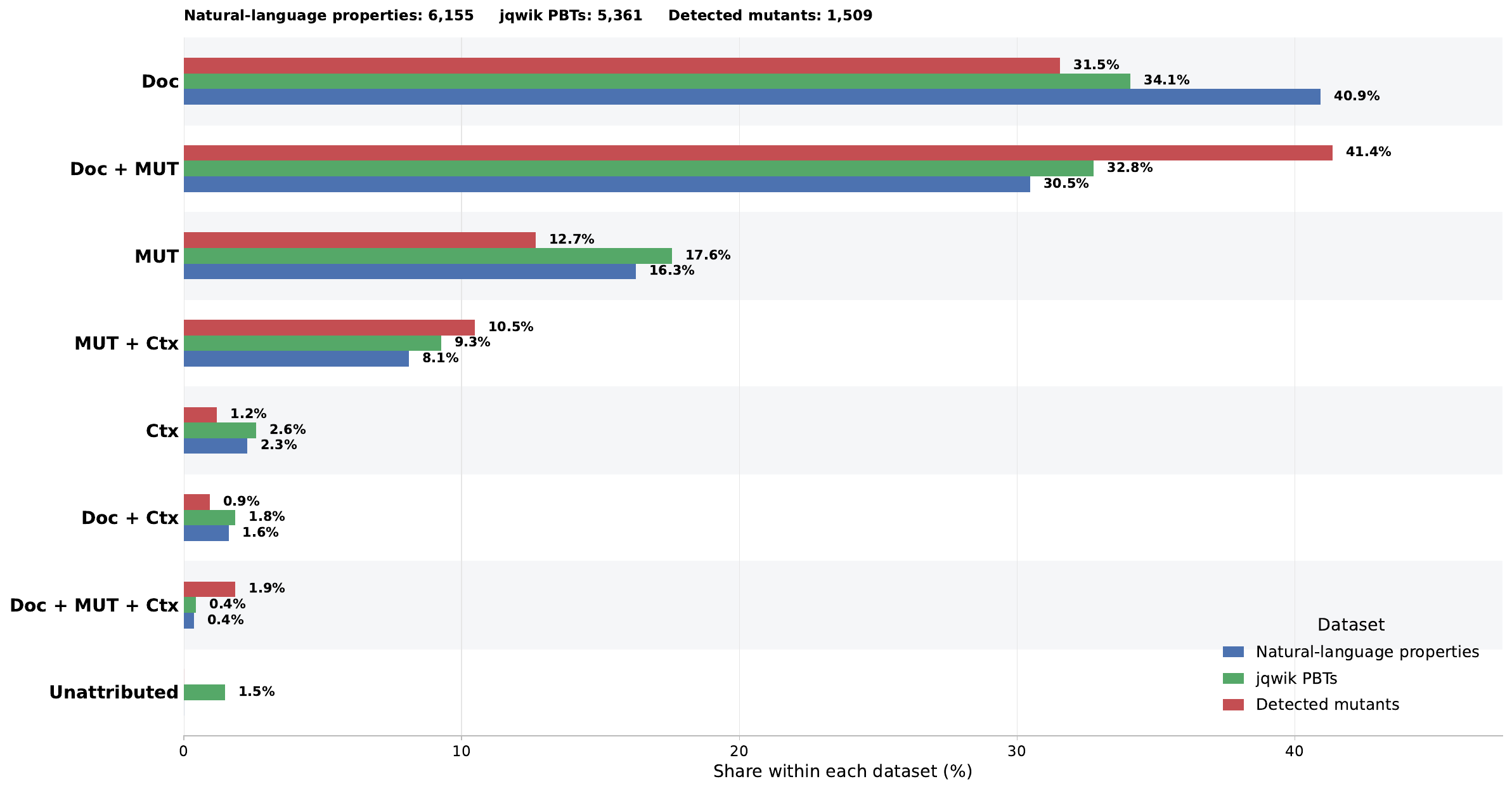}
\caption{Aggregate distribution of the 6{,}155
properties by evidence label, pooled across all configurations
(P1--P8). \textbf{MUT-body}\,=\,focal-method body supplied (P1, P2, P7);
\textbf{MUT-sig}\,=\,signature only (P3--P6, P8); \textbf{Doc}/\textbf{Context}
as in Table~\ref{tab:rq3_context_ablation}. Documentation-derived oracles
dominate overall (Doc 40.9\%, MUT$+$Doc 30.5\% of natural-language properties), while combined-evidence labels concentrate in detection (Doc+MUT rises to 41.4\% of detected mutants).}
\label{fig:source_bar}
\end{figure*}

\begin{table*}[h]
\centering
\caption{Context ablation for property generation.}
\label{tab:rq3_context_ablation}
\scriptsize
\renewcommand\theadfont{\bfseries\scriptsize}
\setlength{\tabcolsep}{2pt}
\renewcommand{\arraystretch}{1}
\resizebox{1.3\columnwidth}{!}{%
\begin{tabular}{@{}cccc@{\hspace{6pt}}ccccc@{\hspace{6pt}}cc@{}}
\toprule
\multicolumn{4}{@{}c}{\textbf{Input}} &
\multicolumn{1}{c}{\textbf{Generated}} &
\multicolumn{2}{c}{\textbf{PBT validity}} &
\multicolumn{2}{c}{\textbf{Baseline}} &
\multicolumn{2}{c}{\textbf{Mutant detected}} \\
\cmidrule(lr){1-4}
\cmidrule(lr){5-5}
\cmidrule(lr){6-7}
\cmidrule(lr){8-9}
\cmidrule(l){10-11}
\thead{Prompt} &
\thead{MUT} &
\thead{Doc} &
\thead{Ctx} &
\thead{\#PBT} &
\thead{Invalid\\(\%)} &
\thead{Valid\\(\%)} &
\thead{Pass\\(\%)} &
\thead{Fail\\(\%)} &
\thead{Killed} &
\thead{Kill\\(\%)} \\
\midrule
P1 & full & \checkmark & all   & 935 & 6  & 94 & 93 & 7  & 328 & \textbf{58} \\
P2          & full & \checkmark & --    & 779 & 13 & 87 & 92          & 8  & 325 & 57 \\
P3          & sig  & \checkmark & all   & 586 & 32 & 68 & 87          & 13 & 103 & 18 \\
P4          & sig  & \checkmark & doc   & 500 & 11 & 89 & 88          & 12 & 224 & 40 \\
P5          & sig  & \checkmark & code  & 857 & 42 & 58 & 87          & 13 & 120 & 21 \\
P6          & sig  & \checkmark & --    & 173 & 10 & 90 & 89          & 11 & 101 & 18 \\
P7          & full & \checkmark & doc   & 992 & 39 & 61 & 87          & 13 & 152 & 27 \\
P8          & sig  & --         & all   & 539 & 10 & 90 & 82          & 18 & 152 & 27 \\
\bottomrule
\end{tabular}}
\vspace{1pt}

\begin{minipage}{0.98\columnwidth}
\scriptsize
\emph{Ctx}: all = class documentation plus caller/callee code and documentation;
doc = class documentation plus caller/callee documentation;
code = class documentation plus caller/callee code;
-- = not supplied.
\emph{Kill} = mutants killed rate.
\end{minipage}
\end{table*}
\vspace{1pt}

\subsection{RQ3: Context Ablation for Property Generation}
\label{sec:RQ3}

RQ3 asks which artifacts in the context bundle help \tool{} generate error-free, intent-driven properties. In order for the property to be error-free, artifacts must include information regarding interfaces available for the property to check. In order for the property to be intent-driven, artifacts must clearly state or imply intended behavior. These qualities matter only if the resulting properties actually expose faults, so for each configuration we also measure how many injected mutants its properties kill. RQ3 therefore investigates how property-generation quality varies across input configurations and how each artifact contributes to property volume, validity, and fault-detection effectiveness.

\paragraph{Experimental Setup.}
We keep the LLM backend, prompts, property-generation pipeline, and \tool{} 
integration fixed, and vary the context bundle to investigate individual contributions of each artifact. We reuse the 240 MUT--Javadoc pairs from RQ1. Table~\ref{tab:rq3_context_ablation} summarizes the eight configurations (P1 to P8) where each row varies the Input bundle, namely the focal method as full body or
signature (\textbf{MUT}) and whether documentation (\textbf{Doc}) and context (\textbf{Ctx}) are supplied. For each prompt configuration, we regenerate \textsc{jqwik} properties using the same two-stage pipeline. Table~\ref{tab:rq3_context_ablation} reports how many properties are generated, how many are valid after synthesis, and how many valid properties pass or fail on the unmutated baseline. It also reports fault detection: following the RQ1 protocol, we run each baseline-valid property against the injected mutants and count how many mutants each configuration kills. Because every configuration draws from the same injected-mutant set, these kill counts are directly comparable across rows. Thus, the ablation measures which artifacts help the LLM produce properties that are grounded, compilable, usable by \tool{}, and fault-revealing. Figure~\ref{fig:source_bar} shows each property's evidence source across the three pipeline stages (natural-language property, jqwik \texttt{@Property}, and mutant detection). These sources are self-reported by the language model at generation and carried through each stage. The causal role of each artifact is established by the ablation itself, which withholds artifacts and measures the effect on validity and kills; Figure~\ref{fig:source_bar} complements this with the model's self-reported provenance, showing that the artifacts the ablation identifies as necessary are also the ones the model reports drawing on.

\paragraph{Results.}
Table~\ref{tab:rq3_context_ablation} shows that the best configuration includes both code and intent. P1 combines the focal method body, its Javadoc, and all surrounding context, producing 935 properties with the lowest invalid rate (6\%), the highest error-free rate (94\%), the highest soundness (93\%), and the most mutants killed (328). This supports the design of \tool{}: executable properties benefit from both semantic intent and concrete program context. Each source is also necessary for the full performance: losing context (P2) increases the error rate from 6\% to 13\%, and losing method under test code (P3) increases the error rate to 32\% — though, as we show below, the body's effect on validity depends on the accompanying context. The two body configurations with full or no context, P1 and P2, dominate detection (328 and 325 kills), well ahead of every signature-only configuration.

Documentation contributes most to generated properties and makes them intent-driven. Figure~\ref{fig:source_bar} shows that documentation contributes towards 73.4\% of generated properties (Doc, Doc + MUT, Doc + Ctx, and Doc + MUT + Ctx), with 40.9\% entirely attributed to documentation. Additionally, the mutant detection results are consistent with these attributions. Documentation contributes towards 75.7\% of mutant detections, with 31.5\% entirely attributed to documentation. This shows  that most properties extracted from the documentation can capture intent and act as an independent oracle.

Detection, however, is bounded by validity. A property that fails to compile can kill no mutant, so a high invalid rate caps how many faults a configuration can find, regardless of how many properties it generates. This resolves why the method body helps in some configurations but hurts in others. The cleanest comparison is P4 and P7, which share documentation-only context and differ only in the focal method. With the signature alone (P4), the model writes conservative properties: 11\% are invalid and 224 mutants are killed. Adding the full body (P7) leads it to write richer properties that call collaborating methods, but documentation-only context does not expose those methods' code, so the calls fail to compile---the invalid rate jumps to 39\% and detection falls to 152, even though P7 generates the most properties (992). The body therefore pays off only when the context can support it: with full context (P1) the collaborator code is present, and with no context (P2) the properties stay focused on the focal method itself, so both reach the highest detection (328 and 325 kills).

The type of context matters more than whether context is present. Holding the signature and Javadoc fixed and varying only the context type, documentation-only context (P4) keeps the invalid rate low (11\%), whereas code-only context (P5) drives it to 42\%. Raw caller/callee code thus adds noise on its own; it becomes useful only alongside documentation, as in the full-context configuration that performs best overall (P1: 6\% invalid, 328 kills). We attribute this to the level at which each source operates: documentation conveys the collaborators' intended behavior and keeps properties at the behavioral level, while code alone supplies low-level detail the model cannot reliably turn into correct properties without that intent.

Documentation and the concrete artifacts play complementary roles. Figure~\ref{fig:source_bar} attributes most behavioral intent to documentation, while the ablation shows that the method body and code context are what make properties compile: removing the body (P3) or supplying it without matching code context (P7) drives the invalid rate up. Documentation states what a property should check, and the body and code context provide the concrete API needed to express it correctly.

\begin{rqanswerbox}{Answer to RQ3}
 Documentation, method under test, and context combine complementary evidence to generate error-free and intent-driven properties, with P1 producing properties with the lowest error rate, and highest soundness. Javadoc supplies intent, playing the biggest role in mutation detection (75.7\%), while method under test and class context supply usage and syntax knowledge improving property error rates.
\end{rqanswerbox}

\subsection{Threats to Validity}

\label{sec:threats}

\paragraph{Construct validity.}
We use source-level injected bugs as controlled current system version faults. This setup directly exercises the oracle problem targeted by \tool{}: regression assertions are derived from the version under test, while intent-derived properties can act as independent semantic oracles. We validate each property on the unmutated baseline before mutation evaluation and discard baseline-violating properties.

\paragraph{Internal validity.}
\tool{} combines LLM-based property generation, context extraction, instrumentation, and evolutionary search. To keep comparisons controlled, we fix the LLM backend, prompt structure, search budgets, and property-generation pipeline across experiments. We also report compilation, integration, and baseline-validation outcomes separately so non-evaluable cases are not counted as bug-detection results.

\paragraph{External validity.}
We evaluate \tool{} on 25 Java systems from the OE25 corpus, including EvoSuite benchmark systems and Apache Commons packages. The current implementation targets Java, JUnit, EvoSuite, and \textsc{jqwik}; however, the core idea---turning executable properties into search objectives---can apply to other testing frameworks that expose property checks and search-guided generation.

\section{Related Work}
\label{sec:related}

\paragraph{Search-based and property-based testing.}
Search-based test generators such as EvoSuite produce high-coverage JUnit suites by evolving object states, primitive values, and method-call sequences, and then adding regression assertions over observed behavior~\cite{Fraser2011EvoSuite,Panichella2018DynaMOSA}. This provides strong \textit{structural reachability}, but the resulting oracles preserve behavior from the current implementation. Property-based testing offers the complementary strength: properties can express semantic expectations independently of a single execution~\cite{Claessen2000QuickCheck,GoldsteinEtAl2024}. However, PBT depends on generators that satisfy selective preconditions; otherwise, inputs are discarded before assertions are reached. Targeted PBT and JQF improve input generation with search or coverage feedback~\cite{LoscherSagonas2017,PadhyeEtAl2019}, but they do not embed executable properties into a DynaMOSA-style generator that constructs complete Java test sequences. \tool{} combines these strengths by turning property assumptions and violations into fitness signals inside search.

\paragraph{LLM-based test and oracle generation.}
Recent LLM-based testing systems improve test generation, assertion generation, and oracle construction, but many remain tied to observed implementation behavior. Execution-driven approaches such as TestChain~\cite{li2024TestChain} capture outputs and treat them as expected results, while coverage- and feedback-driven systems such as CoverUp, TestWeaver, and Cleverest primarily optimize coverage, feedback efficiency, or regression sensitivity~\cite{pizzorno2024coverup,le2025testweaver,liu2025can}. Oracle-generation systems such as TOGA, TOGLL, and Doc2OracLL move closer to semantic correctness by generating assertions or oracles for tests~\cite{togaeval,togll,hossain2025doc2oracll}. However, these systems typically operate on a regression-test prefix: the test sequence is generated first, and the LLM then predicts what should be asserted for that fixed execution. If the prefix reaches buggy current system version behavior, and the oracle is inferred from code-heavy evidence or observed behavior, the generated assertion can still reproduce the implementation rather than challenge it. The oracle remains \textit{post hoc}; it checks a completed test instead of guiding search toward semantically meaningful states.

\paragraph{Positioning.}
\tool{} differs in both timing and mechanism. It generates intent-derived executable \textsc{jqwik} properties from documentation, method code, and caller/callee context, then embeds them directly into DynaMOSA search. Preconditions become progress objectives, property violations become falsification targets, and quantified parameters are bound to values constructed by evolving tests. To the best of our knowledge, \tool{} is the first framework to combine LLM-derived executable properties, property-aware DynaMOSA objectives, precondition-guided fitness, and input materialization in one automated testing loop. This lets testing reason about both \textit{where behavior is reachable} and \textit{whether that behavior violates intended semantics}.

\section{Future Work}

\tool{} establishes executable properties as first-class search objectives, and this foundation opens several extensions. \emph{Precondition-aware generator synthesis} would let the search exploit the structure of \texttt{Assume.that} gates when constructing inputs, reaching properties whose admissible inputs follow strict formats. \emph{Richer property forms}, including relational, metamorphic, and stateful call-sequence would broaden the intent expressible as a search objective beyond single-invocation claims. Finally, the core idea is not Java-specific: evaluation on developer-reported faults and ports to other property engines and languages would test how broadly property-guided search applies.

\section{Conclusion}
\label{sec:conclusion}

This paper presents \tool{}, a property-guided regression-testing framework that unifies \textit{structural reachability} and \textit{semantic falsification}. \tool{} derives intent-driven executable properties from code context, integrates them as first-class objectives in DynaMOSA, and uses property-aware fitness, parameter binding, input materialization, and generator integration to steer search toward both deep states and fault-revealing executions.

Our evaluation on 25 Java systems shows that \tool{} detects current system version bugs missed by regression-derived assertions and reaches hard property preconditions more effectively than standalone property-based testing. These results demonstrate that semantic oracles are most powerful when they guide test generation, rather than merely check completed tests.

\tool{} provides a foundation for future testing systems that ask not only whether software still does \textit{what it did before}, but whether it does \textit{what it should do}. 

\balance
\bibliographystyle{IEEEtran}
\bibliography{main}

\end{document}